# Drug repurposing for SARS-COV-2: A high-throughput molecular docking, molecular dynamics, machine learning, & ab-initio study

Jatin Kashyap*, Dibakar Datta*

Department of Mechanical and Industrial Engineering
New Jersey Institute of Technology
Newark, NJ 07103, USA

*Corresponding authors
Jatin Kashyap, Email: jk435@njit.edu
Dibakar Datta, Email: ddlab@njit.edu

**Abstract**
A micro-molecule of dimension 125nm has caused around 479 Million human infections (80M for the USA) & 6.1 Million human deaths (977,000 for the USA) worldwide and slashed the global economy by US$ 8.5 Trillion over two years period. The only other events in recent history that caused comparative human life loss through direct usage (either by (wo)man or nature, respectively) of structure-property relations of "nano-structures" (either (wo)man-made or nature, respectively) were nuclear bomb attacks of Japanese cities by the USA during World War II and 1918 Flu Pandemic. This molecule is called SARS-CoV-2, which causes a disease known as COVID-19. The high liability cost of the pandemic had incentivized various private, government, and academic entities to work towards finding a cure for these & emerging diseases. As an outcome, multiple vaccine candidates are discovered to avoid the infection in the first place. But so far, there has been no success in finding fully effective therapeutics candidates. In this paper, we attempted to provide multiple therapy candidates based upon a sophisticated multi-scale in-silico framework, which increases the probability of the candidates surviving an in-vivo trial. We have selected a group of ligands from the ZINC database based upon previously partially successful candidates, i.e., Hydroxychloroquine, Lopinavir, Remdesivir, Ritonavir. We have used the following robust framework to screen the ligands; Step-I: high throughput molecular docking, Step-II: molecular dynamics analysis, Step-III: density functional theory analysis. In total, we have analyzed 242,000(ligands)*9(proteins)= 2.178 Million unique protein binding site/ligand combinations. The proteins were selected based on recent experimental studies evaluating potential inhibitor binding sites. Step-I had filtered that number down to 10 ligands/protein based on molecular docking binding energy, further screening down to 2 ligands/protein based on drug-likeness analysis. Additionally, these two ligands per protein were analyzed in Step-II with a molecular dynamic modeling-based RMSD filter of less than 1Å. It finally suggested three ligands (ZINC001176619532, ZINC000517580540, ZINC000952855827) attacking different binding sites of the same protein(7BV2), which were further analyzed in Step III to find the rationale behind comparatively higher ligand efficacy.

1. **INTRODUCTION**

A worldwide global health emergency is posed due to a novel coronavirus(SARS-CoV-2) that is said first to appear and spread from Asia[1]. The World Health Organization (WHO) declared it as Public Health Emergency of International Concern(PHEIC) on 30[th] January 2020[2]. There are similar viruses that emerged in 2003 and 2012, known as severe acute respiratory syndrome coronavirus (SARS-CoV) and middle east respiratory syndrome coronavirus(MERS-CoV), respectively[3]. The new SARS-COV-2 causes a mortality rate of 2%-5% by causing severe acute respiratory syndrome(SARS)[4]. So far, there are known to be seven strains of human coronaviruses, which are broadly classified into two categories, i.e., Alpha coronaviruses and Beta coronavirus[3]. The Alpha coronaviruses are as: 229E, NL63 and Beta coronaviruses are OC43, HKU1, SARS, MERS, SARS-CoV-2. Until the SARS-CoV-2, the SARS and MERS were pathogenic strains with the highest mortality rate of 10% and 36% as per WHO[5]. The causative agent of COVID-19(Coronavirus disease-19) is a positive-sense single-stranded ribonucleic acid (ssRNA) virus encoding multiple proteins[6]. These proteins can be broadly classified into structural and non-structural proteins. The structural proteins include Nucleocapsid(N), Envelope(E), Spike(S), and Membrane(M) proteins. Non-structural proteins include non-structural protein 1-16, i.e., NSP1-NSP16. These proteins play a vital role in the functioning of the SARS-CoV-2, some more than others.

Consequently, some of these proteins are considered a potential target for full or partial inhibition of the SARS-CoV-2[7]. Inhibiting the SARS-CoV-2 by this approach will result in bio-molecules called drugs, and the procedure is known as drug therapy. However, the structures of these proteins can be modified to deactivate their functionality by using 2-dimensional materials-based devices, i.e., MEMS/NEMS[8]. But these approaches are still in the proof-of-concept stage. At the pandemic's beginning, several research efforts resulted in publications citing already proven drugs for other viruses, i.e., Remdesivir, Ivermectin, Chloroquine, Hydroxychloroquine, Lopinavir, Azithromycin, Ritonavir as effective drug therapy for SARS-CoV-2[9]. But their clinical efficacy has been under question in lab trials containing a broader range of patients[10]. That motivated the researchers to reposition other drugs as improved drug therapy with constant efficacy. However, performing in-vivo experiments with new drugs is a very time-consuming and expensive way for drug repositioning. Discovering a new drug altogether can take more than a decade. That is why most in-vivo analyses preferred to focus on in-silico drug repurpose/repositioning before entering the in-vivo phase.

To take it one step further, different in-silico(computational) approaches are advised for drug repositioning/repurposing study[11]. The only constraint is the power of the fastest computer available globally. Agreeably, the results will not be as accurate as those obtained in the in-vivo studies. But the trade-off between speed/money and accuracy justifies the efforts, especially when time can be sped up, i.e., the amount of time it takes to analyze 1000 drugs for repurposing in-silico is a fraction of what it takes for in-vivo studies of the same number of drugs. And the number of drugs in the sample set can reach millions or even billions in in-silico studies which is unheard of in the case of in-vivo studies[12]. Multiple in-silico studies were carried out for the given advantages, targeting one set of proteins in SARS-CoV-2 more than the other.

We started with 242,000 ligands from the ZINC database[13]. The database was selected to include ligands approved by FDA USA and match the LogP and weight of the previously partially successful candidate drugs i.e. Remdesivir, Ivermectin, Chloroquine, Hydroxychloroquine, Lopinavir, Azithromycin, & Ritonavir. Since working with an incorrect binding site even with the correct ligand can severely weaken the approach's efficacy, so we spend rigorous efforts to carefully select the binding sites solely based upon previous experimental studies instead of following an in-silico algorithm-based blind binding-site search. We had identified nine unique binding sites on six different proteins with ZINC ID as follows: 6LU7, 6M0J, 6M71, 6W9C, 6W63, 7BV2 and corresponding binding sites as follows: main protease/inhibitor N3, spike receptor-binding domain, RNA-dependent RNA polymerase, papain-like protease, main protease/inhibitor x77, and nsp12-nsp7/8 complex, Remdesivir & RNA binding sites of SARS-COV-2 system. Based upon docking & ADMET results, we have filtered down the sample set to 18 ligands, i.e., two ligands per binding site. From which, three ligands (ligand $12^{th}$, $13^{th}$, $18^{th}$) were finally selected based upon the results of MD analysis, i.e., less than 1Å RMSD. They all attack the same protein (7BV2) but through different binding sites resulting in multi-binding-site inhibitors. Further, we performed a multi-scale analysis by rationalizing the RMSD and hydrogen bonds formed with the electronic structure analysis obtained from ab-initio approaches. This conclusively proves that the ligand with low RMSD in complex tends to display a direct correlation between their hydrogen bond formations in complex form and the volume covering the frontier molecular orbitals (FMOs) and charge transfer potential of atoms within that region. The same results were corroborated by using Molecular Mechanics Poisson-Boltzmann Surface Area (MMPBSA) derived free energy modeling and machine learning-based FMO energy gap values.

## 2. MODELS AND METHODOLOGY

Our literature survey identified four unique binding sites on protein 7BV2 totaling nine unique binding sites spread over six proteins for the chosen set of 242000 ligands, as shown in Figure S1 in the supplementary document. Table S3 in the supplementary document depicts the relationship between the complex/ligand/protein number used in this work and the complex/ligand/protein's standard identity. Additionally, we have used the same pipeline on Remdesivir in 7BV2 protein as a controlled experiment, as shown in Figure S11. And similar to other complexes, the system seems to be stable from 40 to 100 ns of the production run, and we have considered this part of the trajectory to measure the various MD parameters. This study is broadly divided into three major steps. Step-I performs molecular docking modeling, step-II performs molecular dynamics simulations of the filtered ligands in step-I. Step-III is an electronic structure study of the ligands used in step II and explains the ligands screened out in step-II. It must be highlighted that these three major steps are also associated with minor steps. Before performing step-I, all six proteins were minimized using a molecular dynamics code called YASARA software[14]. YASARA web-server supported by YASARA force fields[15] was used for this minor step and YASARA view for visual inspection[16]. Subsequently, performing high throughput screening-molecular docking of 242,000 ligands on nine different binding sites of 6 unique proteins resulted in more than 2 million distinctive binding-site/ligand combinations using AutoDock Vina software[17].

The top 10 ligands were chosen based upon the least binding energy calculated by the molecular docking method of Step-I. Multiple ligands appear in more than one complex's top 10 ligand list. This proves the robustness of the framework chosen in this study. Subsequently, only two ligands were picked from this list of 10 ligands. The absorption, distribution, metabolism, excretion, toxicity (ADMET) analysis was performed for all ten ligands. The first preference was given to the ligands that show zero violation of LIPSINKI's rules. For non-zero LIPSINKI's rules violations under a given ligand, there were either 1 or 2 violations that were both treated with the same preference, provided that the range from 1st to 10th ligand's binding energy was small for almost all the nine binding sites. And in that case, simply the first and second ligand in order of their binding energy sequence was chosen for the next step, i.e., Step II.

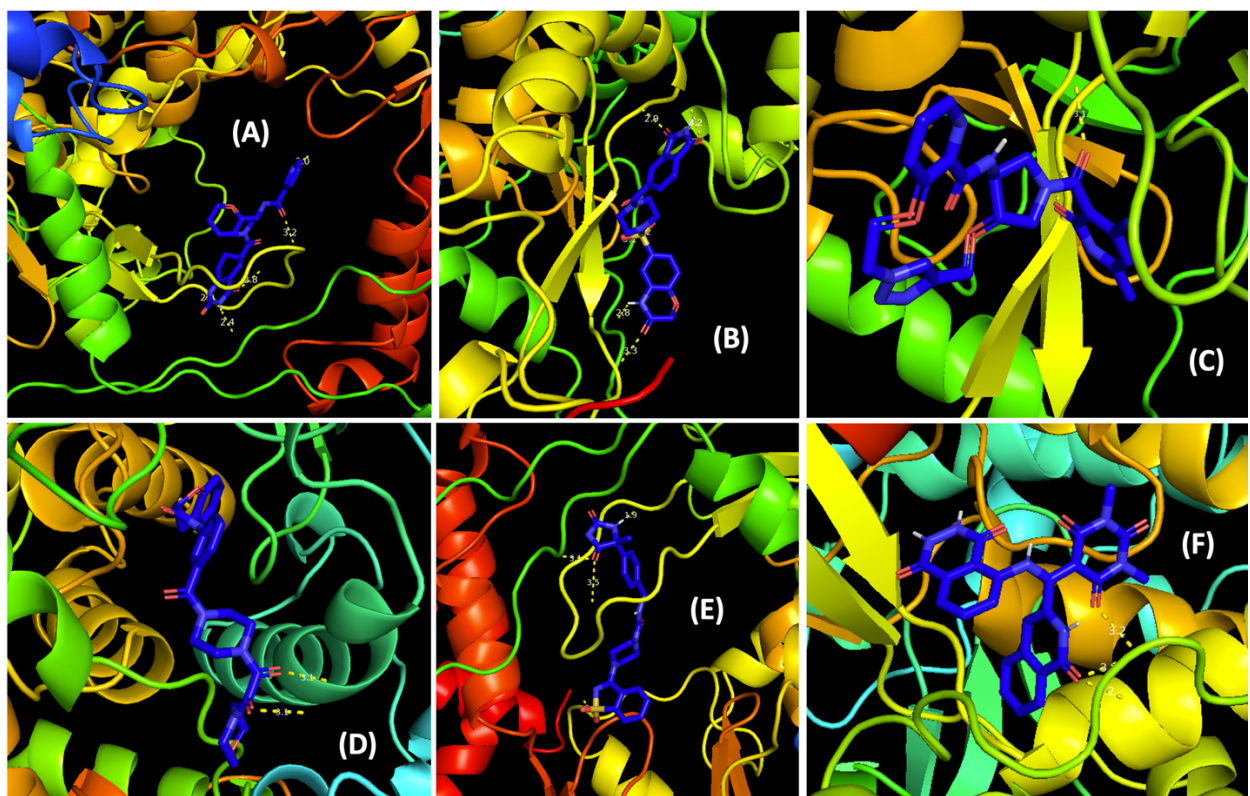

**Figure 1:** Docked complexes of ligands having extreme RMSD. (A,B,C) Lowest RMSD sub-group: 12th, 13th, & 18th ligands, (D, E, F) Highest RMSD sub-group: 5th, 11th, & 14th ligands. Proteins and ligands are shown in cartoon & sticks representation, respectively, with hydrogen bonds highlighted with dashed lines labeled with bonds lengths.

For Step II, molecular dynamics simulations were performed using AMBER20 code[18]. All the complexes were minimized before heating, equilibrating, and finally, a production simulation of 100 ns. In the analysis phase of Step II, root mean square distance (RMSD), root mean square fluctuation (RMSF), hydrogen bond analysis, solvent accessible surface area(SASA) of ligands, and radius of gyration (RoG) of protein & ligand was performed by using the CPPTRAJ[19] utility & python scripts associated with package AMBER20. Furthermore, MMPBSA.py subroutine was used to perform Molecular Mechanics Poisson-Boltzmann Surface Area (MMPBSA) free energy analysis of the complexes. We have selected three ligands with the highest RMSD (high RMSD subset) and three ligands with the lowest RMSD (low RMSD subset) for further in detail investigation, as shown in Figure 1. In step III, VASP[20] was used to perform the following analysis: Bader charge[21], Highest occupied molecular orbital(HOMO), lowest occupied molecular orbital(LUMO), molecular electrostatic map(MEP). In addition, quantum chemical characteristics were calculated, which are various functions of HOMO & LUMO energies. Most of the analysis and discussion is focused on the high RMSD and low RMSD ligand subset while discussing the rest of the complexes when necessary.

## 2.1 Ligands and Proteins Preparation

We have downloaded all six protein structures from the RCSB protein data bank[22]. After downloading the structures, they were processed in the UCSF Chimera package[23]. All hetero-atoms were removed, i.e., co-crystallized ligands, ions, water molecules. In the case of multi-binding site protein 7BV2 protein, we removed the co-crystallized ligands/molecules based upon the binding site. For example, we pulled co-crystallized RNA strands to use the binding site of the RNA strand as the binding site of the potential ligand. We kept the RNA strand intact in the case of using the interface of NSP12/7 or NSP 12/8 as the binding site of the potential ligand. There is one exception to the framework of removing one co-crystallized molecule while keeping the other. Since this protein has Remdesivir co-crystallized, it was trimmed in case of using any four of the binding sites available on the protein for the potential ligand. This exception was made because out of four binding sites available on the protein 7BV2, only one binding site has the co-crystallized molecule, which works as an inhibitor, i.e., Remdesivir.

To avoid the existence of the dual molecules crystalized with the protein, which can act as an inhibitor, we removed Remdesivir from all four "allotropes" of protein 7BV2. Since the proteins had gone through trimming, which may lead to dangling bonds or steric clashes between the chains, they were minimized using the conjugate gradient method implemented in YASARA. The proteins were minimized until the YASARA's Z score was achieved within an acceptable tolerance for all nine binding sites. Consequently, a thorough literature survey was conducted to find the binding sites of the proteins from experimental studies all over the world. Following the proteins minimization, a Linux bash shell script was developed to be used on the XSEDE supercomputing machine[24] to execute high throughput virtual screening in step-I, as described above. The ligands were not minimized before high-throughput screening since their structure was not modified as in the case of proteins. This drastically reduces our computational requirement and makes it feasible to perform molecular docking on 242,000*9= 2.178 million unique combinations of a protein-binding site and a ligand.

## 2.2 Molecular Dynamics

After completion of Step-I, two ligands were picked for all nine proteins. Those 18 combinations were processed through Step-II, i.e., Molecular Dynamics simulation. We used AMBER20 software for the MD simulations. The tasks performed in step II were minimization, heating(1ns),

equilibration(10ns), production run(100ns), followed by post-production analysis. For the minimization run, we performed 10,000 steps. First, we computed 5000 steps of the steepest descent method followed by 5000 steps of the conjugate gradient method. For heating, we raised the temperature in a stepwise manner. In the first step, the temperature was increased from 0°C to 100°C during 0.3 ns followed by 100°C to 200°C during the next 0.3 ns before throttling it to the final 300°C for the remaining 0.4 ns. The last leg contains additional 0.1 ns compared to the 1$^{st}$ and 2$^{nd}$ leg of 0.3 ns to compensate temperature rise from a point which was highest among all beginning temperatures among all three legs despite being raised by an equal amount, i.e., 100°C. A temperature coupling constant of 1.0ps was used for the heat bath for all legs. We used the NVT ensemble for this part of the MD step. For the equilibrium step, the system was equilibrated for 10ns with an NPT ensemble. This step helps stabilize the system while relaxing any compressed/stretched bonds raised due to the last stage of heating. Finally, the production run was implemented for 100ns with the NPT ensemble. Following settings applied for all MD simulations steps: SHAKE algorithm to constrain hydrogen bonds, step size of 2 fs, cutoff of 9.0 Å, restrain weight of 0.5. For MD analysis, we wrote every 1000$^{th}$ frame to the trajectory file output of the production run of 50 million steps resulting in the trajectory file containing 50,000 frames. We performed the RMSD, RoG, SASA on every frame of the trajectory file and every 100$^{th}$ frame for MMPBSA analysis. One frame should be considered equivalent to 2 fs for hydrogen bond analysis in this work.

## 2.3 Density functional theory

We used the VASP program to optimize the ground state geometries of all the 18 ligands before performing the ab-initio analysis. The density functional theory-based program used Becke3-Lee-Yang-Parr(B3-LYP) exchange functional, Lee-Yang-Parr (LYP) correlation functional with standard 6-311G basis set. The Koopman theorem was used for plotting the HOMO and LUMO of the ligands. Bader charge program was used to perform the atomic charge analysis[25]. Which was substituted for MEP maps highlighting the electron-rich and deficient areas at various points in the region surrounding the molecule and at equidistance from the molecular surface.

## 2.4 Drug Likeness Analysis

We used the SwissADME online web-server [26] to perform the drug-likeness analysis. Drug-likeness helps find the probability of a drug working as an oral consumption drug based upon its bioavailability. It is solely based upon the physicochemical and structural characteristics of the compounds. It is general practice in the pipeline for computer-aided drug design to screen the compounds according to pharmacokinetics. The SwissADME server gives access to five filters for drug-likeness screening based upon five unique methodologies to filter out the drugs based upon pharmacokinetics. Pharmaceutical companies mainly invented these filter methods to better their drug discovery pipelines. The five rules are Lipinski, Ghose, Veber, Egan, and Muegge; the corresponding inventor-corporation are Pfizer, Amgen, GSK, Pharmacia, and Bayer, respectively. Although all five filters can be used to find out the best combination for the given compounds, we primarily used Lipinski's rule of five. Since Lipinski's rule is the oldest among all five filters, it is most acceptable among the computational biology research community.

We considered Lipinski's rule as the primary filter and the remaining four filters as secondary collectively. Although we also measured the parameters from the rest of the four filters, their corresponding decision-making weight was less than Lipinski's rule. Lipinski's rule was over-ridden only when we encountered a severe breach of secondary filter for a given compound. In addition to the drug-likeness screener, we considered the following screeners supported by the same web server: Physicochemical properties, Lipophilicity, water-solubility, Pharmacokinetics, medicinal chemistry. Out of these, only drug-likeness was considered as the primary screener. And further down, the Lipinski rule was regarded as the primary drug-likeness filter. Lipinski rule evaluates a compound based on pharmacokinetics, absorption, distribution, metabolism, excretion. It has the following selection criteria: molecular mass is less than 500 daltons, octanol-water partition coefficient(logP) does not exceed 5, hydrogen bond donors(HBD) is less than 5, hydrogen bond acceptors(HBA) is not more than 10.

**2.5 Machine Learning**

We used GNN (graphical neural network) with two layers based upon Battaglia equations [27] using an applied ML book[28]. These equations are mostly used for bio-molecules training. The model was trained on a dataset known as QM9[29] containing 134,000 molecules/ligands/inhibitors/compounds. This neural network converts the distance between the atoms to an inverse pairwise distance. This inverse pairwise distance acts like the edges of the "graph" in Graphical Neural Network. It fits well with the framework that more weightage is assigned to the atomic pairs having less distance than atomic pairs having comparatively more distance to mimic the conditions of the more interaction between the atoms involved in the former case than the latter. Moreover, the limit conditions need to be satisfied, i.e., the weights shall never be zero and infinity. The model is designed so that the inverse pairwise distance cannot be zero and infinity. But our model is tweaked to assign zero weights if the pairwise inverse distance crosses a threshold. That is equivalent to assigning a cutoff value for force-field-based classical molecular dynamics approaches[30][31][32][33].

## 3. RESULTS AND DISCUSSION

In the following in-depth discussion, Table 1 can be used as a reference aid in understanding the dependability among different parameters investigated in this work and discussed in the upcoming sections.

**Table 1:** Proportionality of different investigated parameters with RMSD.

| Property | Relationship with RMSD |
|---|---|
| *DFT parameters* | |
| HOMO-LUMO gap (reactivity & stability) | Directly proportional |
| Ita(hardness) | Directly proportional |
| Omega(electrophilicity) | Inversely proportional |
| Sigma (local softness) | Inversely proportional |
| Zita(electronegativity) | Inversely proportional |
| *MD Parameters* | |
| Protein RoG | Directly proportional |
| Ligand RoG | Inversely proportional |
| SASA | Inversely proportional |
| Binding free energy of complexes | Directly proportional |

### 3.1 Drug Likeness Analysis

We passed the top ten ligands sorted according to the binding energy obtained from the molecular docking analysis from each of the nine binding sites of six proteins through the SwissADME web-server[26] and then picked the two ligands from each of the binding sites. The selection criteria were the binding energy and violation of Lipinski's rules. We made the following observations during this sub-step: 1) a violation of Lipinski's rule (if any) is not more than one; 2) only one protein(6M0J) does not have at least one ligand with zero Lipinski's violation; 3) two proteins (7bv2_nsp12-7 & 7bv2_rna) with only one ligand with zero Lipinski's rule violation. Except for the above-highlighted cases, every other case has at least two or more ligands with zero Lipinski's rule violations. This accounts for the efficacy of the drug discovery pipeline chosen in this work. Furthermore, in multiple cases, the same drug demonstrated good binding potential to various proteins in a selected set of nine unique binding sites from six proteins. For example, the compound ZINC000750965621 is present in the top 10 best binding ligands list of protein 6M0J & 7BV2_rna. Similarly, drug ZINC000587983851 is shown up in the list of 6M0J & 6M71_nsp_12_7_8. Such drugs may have a higher potential to be selected in the final pool of drugs to be synthesized, given they can target more than one binding site, increasing their inhibition ability. But in our cases, the binding energy of both the above drugs was less than the binding energy of other drugs having 0 Lipinski's rule violation, so we could not select these drugs for those proteins/binding sites, given the already specified rules.

## 3.2 Molecular Dynamics Study

### 3.2.1 RMSD and RMSF Analyses

#### 3.2.1.1 RMSD

Following three ligands exhibit the least RMSD (<1Å) - Ligand $12^{th}$, $13^{th}$, $18^{th}$ i.e. 7BV2_nsp12-7/ZINC001176619532, 7BV2_nsp12-8/ZINC000517580540, 7BV2_rna/ZINC000952 855827 as 0.6Å, 0.45Å, and 0.5Å, respectively. Since all the sub-1Å RMSD ligands docked to the same protein 7BV2 but through different binding sites, future in-vivo tests shall be done in this protein, at least for selected ligands in this work. The HETAOMS removed in the first, second, and third cases are interaction sites of NSP12-7 interface (7BV2_nsp12-7), NSP12-8 interface(7BV2_nsp12-8), and template-primer RNA(7BV2_rna). Figure S2 in the supplementary depicts the RMSD analysis of the rest of the complexes. Similarly, we selected a set of 3 ligands with the highest RMSD to juxtapose against the earlier set of least RMSD ligands. These ligands are $5^{th}$, $11^{th}$, $14^{th,}$ i.e., 6m71_nsp12_7_8 /ZINC000410177506, 7bv2_nsp12-7/ZINC000616537204, and 7bv2_nsp12-8/ZINC001180048431 with RMSD as follows: 4.1Å, 2.05Å, and 1.9Å. In this work, we had primarily discussed hydrogen bonding among all bonded interactions. However, we performed calculations considering all kinds of interaction a biological system undergoes, i.e., van der Waal, Coulomb interactions in MMPBSA analysis, etc.

We observed that a ligand tends to have lower RMSD if it: a) forms a greater number of hydrogen bonds with the binding site residues and b) maintains those bonds for longer simulation time. Among least RMSD sub-group, ligand 12 have RMSD of 0.6Å primarily because of stability acquired by forming a hydrogen bond with residues PHE362 through terminal pyrimidine as shown in $1^{st}$ circle and ALA467 & ALA470 through hexahydroindan & double-bonded oxygen as shown in $1^{st}$ and $2^{nd}$ dip in $2^{nd}$ circle in Figure 3(a). Correspondingly, very high charge transfer values (Figure 8A) and HOMO/LUMO orbitals iso-surfaces (Figure 10A) over ligand can be seen in the region containing these residues, i.e., PHE362 and ALA470. The second ligand in the least RMSD sub-group is ligand 13, with an RMSD of 0.45Å. It forms hydrogen bonds with ALA478, THR485, ASN417, ASN 427, GLN461, ASN463. All these residues were encapsulated using a single closed curve in Figure 3(B). Out of all hydrogen-bonded residues, only ASN 427, GLN 461, ALA 478, ASN463 are covered by the iso-surfaces of HOMO/LUMO and extreme charge transfer values as shown in black closed curves in Figure 8(B) and Figure 10(B), respectively. The last ligand of the least RMSD sub-group is ligand 18, with an RMSD of 0.5Å. This ligand forms hydrogen bonds with THR485, ARG489, SER484 (first circle in Figure 3c), and ALA60 (highlighted by the second circle in Figure 3c). Furthermore, we observe a direct correlation of these hydrogen bonds formation with ab initio analysis (Figure 8c and Figure 10c), which further proves the rationale of these bonded interactions for the lower RMSD of these ligands.

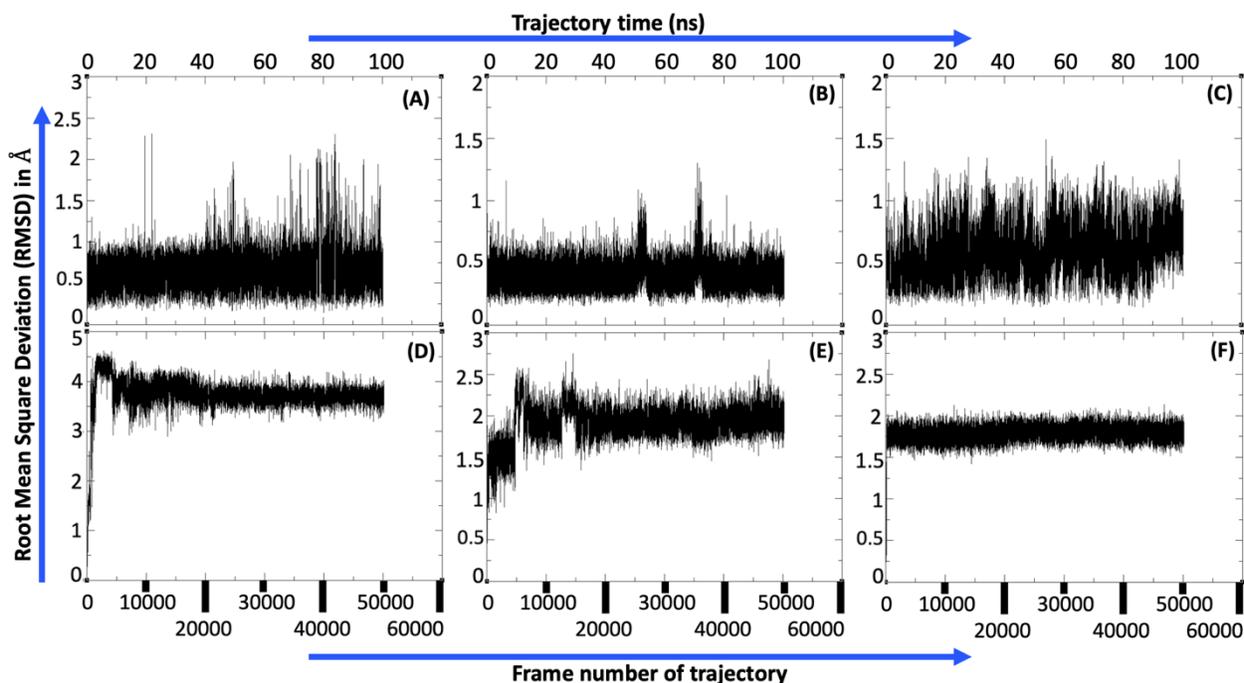

**Figure 2:** Root Mean Square Deviation (RMSD) of ligands having extreme RMSD. (A, B, C) Lowest RMSD sub-group: 12th, 13th, 18th ligands. (D, E, F) Highest RMSD sub-group: 5th, 11th, 14th ligands. The X-axis shows the frame number(bottom) & trajectory time(top) of the production run, and the y-axis is RMSD (Å).

Among the high RMSD ligand sub-group, ligand 5 has an RMSD of 4Å and significantly fewer bonded interactions with the protein binding site. The ligand has two sites for hydrogen bonds, i.e., ARG291 and PHE338. Both bonds are neither stable nor high frequent bonds (Table 1a). It can be attributed to comparatively less HOMO/LUMO iso-surfaces volume and lower extreme charge transfer values, as highlighted by closed black curves superimposed on Figure 8D and Figure 10D. The next ligand in the high RMSD group is ligand 11, with an RMSD of 2Å. This ligand has the largest number of hydrogen bond sites, where the bond exists for more than 1000 frames. It is ranked number 2 (according to RMSD) in the high RMSD ligand sub-group. We believe this complex exhibits this behavior because although the number of 1000s frames with hydrogen bond sites is five times the other ligands, it cannot retain those hydrogen bonds for an extended period. For example, the longest hydrogen bond in this ligand exists for 5634 frames (Table 1a), but this number is 30,000 frames on average (Table1b) for the ligands in the low RMSD sub-group. This seems to be the primary reason for the high RMSD of this ligand in the complex.

The last ligand from the high RMSD sub-group is ligand 14 with 1.9Å RMSD. This ligand has every profile of a ligand belonging to the low RMSD sub-group, because of which it has the least RMSD among its sub-group. But there is one difference in terms of the HOMO/LUMO iso-surfaces region. Opposite to a ligand from the low RMSD sub-group, the HOMO/LUMO iso-surfaces of this ligand are not overlapping, as shown in Figure 10(F). And this makes it comparatively less stable in the binding site. The HOMO iso-surfaces cover the extreme right N-N bond edge of tetrahydro-3,6-pyridazinedione of 3,4-Diazabicyclo[4.4.0]decane-2,5-dione terminal, and LUMO iso-surfaces covers the 2,4(1H,3H)-Pyrimidinedione, dihydro-1,3-dimethyl attached in the middle of the ligand chain between azadecalinone and tetrahydro-3,6-pyridazinedione of 3,4-Diazabicyclo[4.4.0]decane-2,5-dione.

### 3.2.1.2 *RMSF*

We have focused on the RMSF of only those residues within 5 Å distance from the ligands. Those residues were identified and visually inspected (Figure S10) by PyMol, and investigated with the RMSF plot of that complex as generated by AMBER20's CPPTRAJ utility. For Ligand 12, the residues selected are VAL 330, HIS 359, PHE 360, PHE 361, PHE 362, ALA 363, GLN 364, LYS 465, TYR 466, ALA 467, ILE 468, SER 469, ALA 470, LYS 471, ASN 472, ARG 473, ALA 474, ARG 475, GLU 731, SER 734, ARG 756. All residues belong to main chain A of the protein.

For complex ligand 13, the selected residues are VAL 415, ASN 417, LEU 418, LYS 420, SER 421, ALA 422, ALA 432, TYR 436, GLN 461, MET 462, ASN 463, LEU 464, LYS 465, TYR 466, VAL 477, ALA 478, GLY 479, VAL 480, SER 481, ILE 482, THR 485, ARG 489, SER 602, GLY 603. All residues belong to main chain A of the protein.

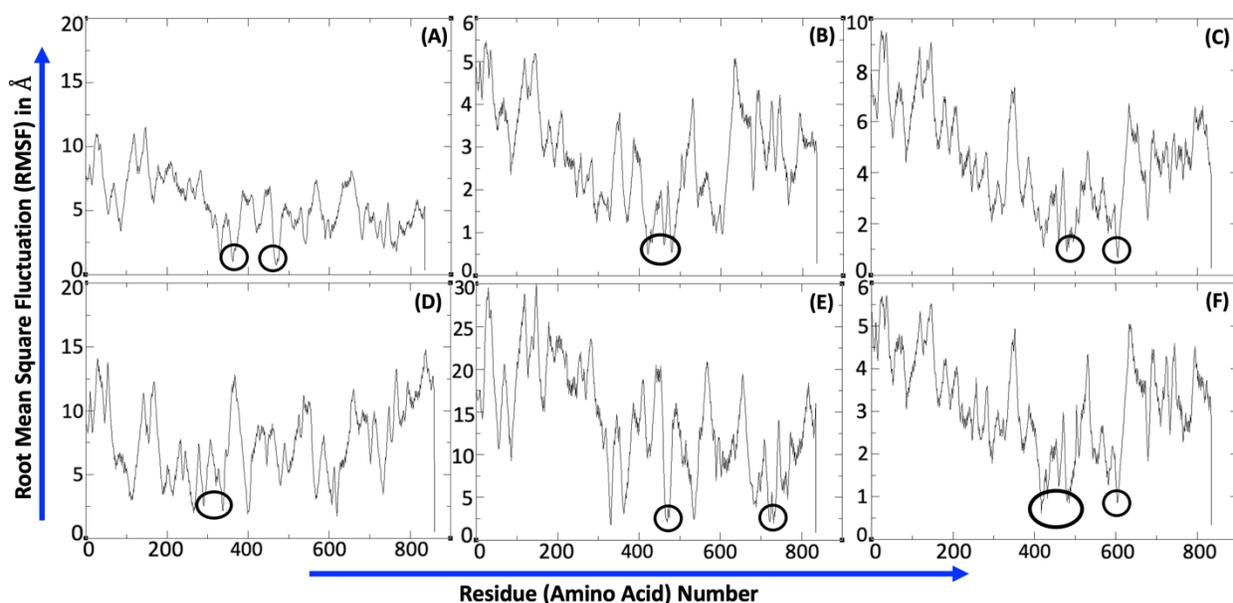

**Figure 3:** Root mean square fluctuation of ligands having extreme RMSD. (A,B,C) Lowest RMSD sub-group: 12th, 13th, 18th ligands. (D, E, F) Highest RMSD sub-group: 5th, 11th, 14th ligands. The X-axis shows the residue identification, and the y-axis is RMSF (Å).

For complex ligand 18, the following residues are selected- LYS 420, SER 421, ALA 478, GLY 479, VAL 480, THR 485, ARG 489, GLN 493, LEU 496, SER 602, GLY 603, ASP 604, ALA 605, ALA 608, TYR 609. All residues belong to main chain A of the protein. Figure S3 in the supplementary depicts the RMSF plots of the rest of the complexes.

The above-selected residues having hydrogen bond with ligands shows the least RMSF within corresponding selected ligand-protein complexes, i.e., ligand 12, 13 & 18 with the RMSF of 5.1 Å,

1.5 Å, and 2.2 Å. Furthermore, Figure 3 shows the RMSF plot for all low and high RMSD subset ligands with superimposing black closed curves highlighting the residues forming the highest number of hydrogen bonds with the protein as determined by hydrogen bond analysis (described in next section). Among the low RMSD subset, the ligand 12 forms hydrogen bond for 2510 and 26418 frames with the N and O atoms of residue 2-Amino-3-phenylpropanal (PHE 362), respectively. Ligand 13[th] forms hydrogen bond for 679, 18673, 32391 frames with residues 3-Amino-4-oxobutanamide (ASN417), 2-Amino-3-oxobutanal (THR 485) and Propanal, 2-amino- (ALA 478) residues, respectively. Ligand 18[th] maintained hydrogen bonds for 25023 and 29790 frames with Propanal, 2-amino- (ALA 605), and 2-Amino-3-hydroxybutanal (THR 485) residues.

Among the high RMSD subset, ligand 5[th] has 6 and 8 hydrogen bonds with 2-Amino-3-phenylpropanal (PHE 338) and 2-(4-Amino-5-oxopentyl)guanidine (ARG 291) residues, respectively. Ligand 11[th] holds hydrogen bonds for 4031 and 15573 frames with Propanal, 2-amino- (ALA 467), and 2,6-Diaminohexanal (LYS 718) residues, respectively. Ligand 14[th] forming hydrogen bonds for 7510, 23435, 32130 frames with 2-Amino-3-(4-hydroxyphenyl) propanal (TYR 436), Propanal, 2-amino- (ALA 478), 2-Aminoacetaldehyde (GLY 603), respectively. The above information is used from the next section, and it shows justification of the RMSF characterization. Low RMSF residues have a relatively higher number of hydrogen bonds for stability, resulting in lower fluctuation than other residues of the protein.

### 3.2.2 *Hydrogen Bond Analyses*

We calculated the hydrogen bonds formed between the ligand and the protein using the hydrogen bond CPPTRAJ tool in the AMBER20 package. Consequently, we sorted out the output by the number of hydrogen bonds formed. Then we trimmed the data to show the highest among the following in Table 2a and 2b: (i) bottom-most five lines or (ii) the number of lines having hydrogen bonds existing over 1000 frames or (iii) the number of lines having the highest lifespan of a hydrogen bond in that complex to be visible in the selection.

| Table 2a: Calculated hydrogen bond statistics for high RMSD ligand subset. | | |
|---|---|---|
| **Longest Lifespan** | **Total Frames** | **Atoms/Residues Identity** |
| **5[th] Ligand** | | |
| 1 | 6 | LIG_860@O4-PHE_338@N-H |
| 1 | 8 | LIG_860@O3-ARG_291@NH2-HH21 |
| **11[th] ligand** | | |
| 132 | 1164 | LIG_835@O2-SER_734@OG-HG |
| 26 | 1235 | LIG_835@O2-LYS_718@NZ-HZ3 |
| 30 | 1361 | LIG_835@O2-LYS_718@NZ-HZ2 |
| 3 | 1612 | SER_469@OG-LIG_835@N-H |
| 9 | 2201 | LIG_835@O3-PHE_362@N-H |
| 10 | 3260 | ILE_468@O-LIG_835@N-H |

| 29 | 4031 | ALA_467@O-LIG_835@N4-H1 |
| 28 | 4671 | LIG_835@O1-LYS_718@NZ-HZ1 |
| 30 | 5268 | LIG_835@O1-LYS_718@NZ-HZ2 |
| 32 | 5634 | LIG_835@O1-LYS_718@NZ-HZ3 |
| **14th Ligand** | | |
| 6 | 2590 | LIG_835@O1-ASN_417@ND2-HD22 |
| 7 | 6132 | ASN_417@O-LIG_835@N2-H |
| 6 | 6593 | LIG_835@O2-GLN_493@NE2-HE21 |
| 10 | 7510 | LIG_835@O3-TYR_436@OH-HH |
| 23 | 23435 | ALA_478@O-LIG_835@N4-H2 |
| 25 | 32130 | GLY_603@O-LIG_835@N5-H3 |

**Table 2b:** Calculated hydrogen bond statistics for low RMSD ligand subset.

| Longest Lifespan | Total Frames | Atoms/Residues Identity |
|---|---|---|
| **12th Ligand** | | |
| 1 | 26 | PHE_362@N-LIG_835@N1-H |
| 2 | 108 | LIG_835@O4-ALA_470@N-H |
| 2 | 257 | LIG_835@N2-ALA_467@N-H |
| 4 | 2510 | LIG_835@O3-PHE_362@N-H |
| 19 | 26418 | PHE_362@O-LIG_835@N1-H |
| **13th Ligand** | | |
| 13 | 319 | LIG_835@O4-ASN_463@ND2-HD22 |
| 28 | 442 | LIG_835@O3-ASN_463@ND2-HD22 |
| 8 | 570 | LIG_835@O4-GLN_461@NE2-HE22 |
| 3 | 616 | LIG_835@O3-ASN_427@ND2-HD21 |
| 3 | 679 | LIG_835@O2-ASN_417@ND2-HD21 |
| 13 | 18673 | THR_485@OG1-LIG_835@N2-H |
| 20 | 32391 | ALA_478@O-LIG_835@N3-H1 |
| **18th Ligand** | | |
| 3 | 668 | LIG_835@O4-SER_484@OG-HG |
| 5 | 1060 | LIG_835@O2-ARG_489@NH1-HH11 |
| 3 | 2340 | LIG_835@N6-THR_485@OG1-HG1 |
| 18 | 25023 | ALA_605@O-LIG_835@N1-H |
| 28 | 27450 | LIG_835@O4-THR_485@OG1-HG1 |

Table 2 displays hydrogen bond data among three columns. The first column is maximum lifespan, indicating the number of consecutive frames a hydrogen bond exists without breaking up between a given protein and ligand. The second column is the total number of frames the hydrogen bond exists between a given protein and ligand despite being non-consecutive frames. The third columns provide information about the exact identity of atoms involved in the hydrogen bond formation from protein and ligand structure, i.e., donors, acceptors, the hydrogen atom. The data from the first two columns in Table 2 support the results obtained in the RMSD section. To restate, the low RMSD sub-group has ligands 12th, 13th, 18th with 0.6 Å, 0.45 Å, and 0.5 Å RMSD, respectively. We have obtained almost the same order from the first two columns of hydrogen bond formation [Table2b]. For the following analysis, lines of a table shall be read in reference to Table(2a,b). In the 13th complex, due to the 1,2-Cyclohexanedicarboximide and Piperazine, the bottom two lines (Table1b) have 18673 & 32391 frames with 13 & 20 frames of maximum lifespan. Due to the Piperidin-3-one & carbonyl group in the 18th complex, the bottom two lines have 25023 & 27450 frames with 18 & 28 frames of maximum lifespan, respectively. Lastly, for the 12th complex, the bottom two lines have 2510 & 26418 frames with a maximum lifespan of 4 & 19 frames, respectively, through 4a,5,6,7,8,8a-hexahydro-1H-quinazoline-2,4-dione in the ligand. In addition to the number of frames for hydrogen bond existence, it is also essential to know the maximum amount of period, in terms of frames, that the same bond did not break.

For complexes 13th and 18th, both columns have a high number for the bottom two lines in Table[2b] than the complex 12. After sorting, if the bottom two lines in Table[2a,2b] show fewer frames for a given hydrogen bond and the lesser maximum lifespan, that will reflect in the RMSD value of the ligand compared to protein. The former will be inversely proportional to the latter. Therefore, we have a high number in the bottom two rows for the 13th and 18th complex but a lower number for the 12th complex since the 12th ligand has the maximum RMSD among the subset of lowest RMSD cases. An almost similar correlation was observed in the high RMSD ligand subset. The RMSD of the 5th, 11th, 14th ligands is 4 Å, 2 Å, and 1.9 Å, respectively. For the 5th ligand, carbonyl group & hexahydroindan contributed to the bottom two lines in Table [2a] to have 6 & 8 frames with 1 & 1 frames of maximum lifespan. For the 11th complex, the bottom two lines are 5268 & 5634 frames, with 30 & 32 frames as maximum lifespan, respectively, due to Sulphur attached hexahydroindan. For the 3,4-Diazabicyclo[4.4.0]decane-2,5-dione in the 14th complex, the bottom two lines are 23435 & 32130 frames, with 23 & 25 frames as maximum lifespan, respectively. Hence, the combination of the number of frames a hydrogen bond exists and maximum lifespan correlates well with the ligand RMSD observed for that complex in Figure 3. Table S1 in the supplementary enlists the hydrogen bond analysis of the rest of the complexes.

### 3.2.3 *Radius of Gyration (RoG)*

RoG measures the spread of the mass of a given molecule around its geometrical central axis. Comparing the RoG plot of a ligand and the corresponding docked protein reveals two aspects: the average value's relative difference and the fluctuation range's difference. For the former, the ligand RoG has values around 4-6 Å, but for the latter, the RoG ranges between 39 Å to 51 Å among all the 18 complexes considered. On average, the fluctuation range for the ligand and protein RoG are about 0.2 Å and 0.07 Å, respectively.

*3.2.3.1 RoG of ligands*

A ligand with RMSD less than 1 Å helps the docked protein maintain its RoG, i.e., less fluctuation. But it will have less impact on decreasing the RoG of a protein if it already has high RoG in the first place, to begin with. To elaborate, the 2 out of 3 given complexes with the highest, i.e., 5[th], 11[th], and 14[th] complexes, are showing such behavior. We can observe sudden jumps in RoG values from 0-2000th frame (6.2 Å to 4 Å) for the 5[th] complex (Figure 4d) and then multiple jumps on the RoG plot for the 11[th] complex between 5000th and 15000th frames with the range of 6.4 Å to 5.4 Å (Figure 4e). We considered the 14[th] complex as an outlier. Despite having high RMSD, it does not show such sudden jumps on the RoG plot.

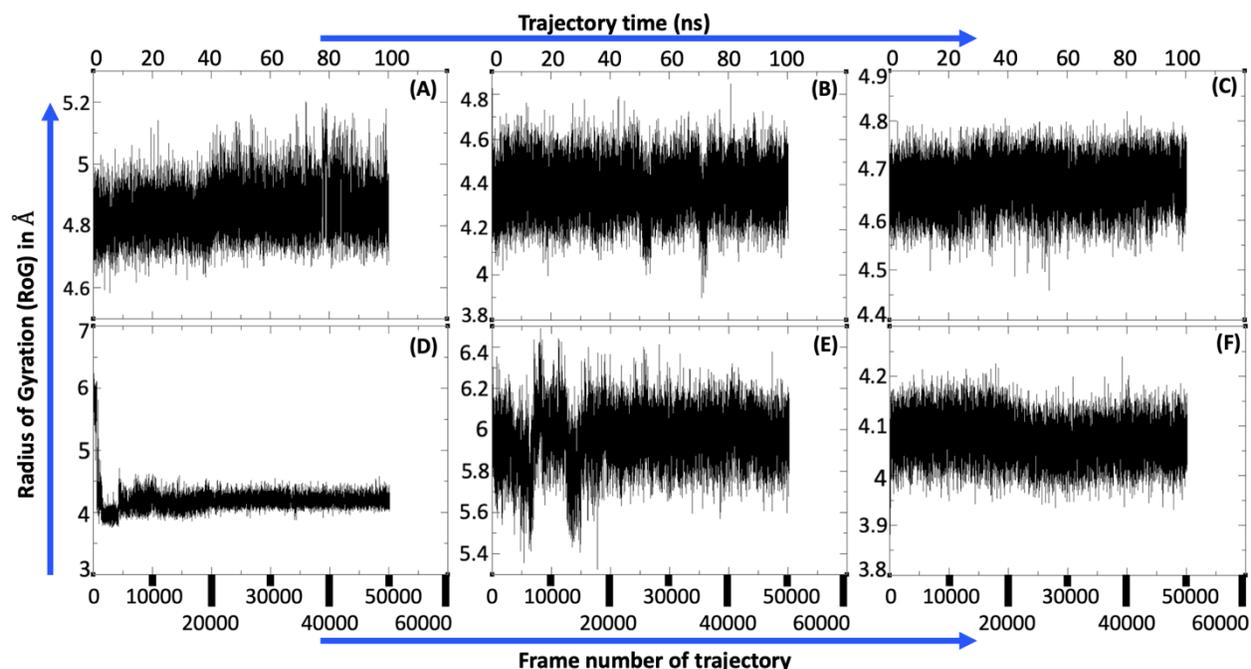

**Figure 4:** Radius of Gyration (RoG) of ligands having extreme RMSD. (A,B,C) Lowest RMSD sub-group: 12th, 13th, 18th ligands. (D, E, F) Highest RMSD sub-group: 5th, 11th, 14th ligands. The X-axis shows the frame number(bottom) & trajectory time(top) of the production run, and the y-axis is the ligand's RoG (Å).

The ligands with the least RMSD, i.e., 12[th], 13[th], 18[th], show almost opposite behavior to those with higher RMSD ligands. There are no sudden jumps on the ligand RoG plots (Figure 4a, b, c) except a slight fluctuation for the 13[th] complex (Figure 4b). However, we can consider it an outlier since the fluctuation only stays from 36000th to 37000[th] frames. Overall, for the three complexes with the lowest RMSD, the ligand RoG varies between 4.3 Å to 4.8 Å without any significant change in the RoG, i.e., step jump on the ligand RoG plot, during the production simulation. That indicates that the complexes' geometry remains stable, i.e., no drastic change in the distribution of the masses around the central axis. Furthermore, no drastic change in the complex's structure means no significant switch in the function of the complex since for biomolecules, their folding/unfolding dictates their behavior more than the atom constituting the biomolecule[34]. Figure S4 in the supplementary depicts the RoG plots of the rest of the ligands.

*3.2.3.2 RoG of protein*

We believe for proteins as well, it is more important to have a system with less RoG variation. The RoG of all 18 proteins has average values of around 50 Å (Figure 5), which is higher than the ligands. Therefore, we can follow the hypothesis mentioned above in all 6-complexes considered (3 high RMSD subset ligands+ 3 low RMSD subset ligands). Figure S5 in the supplementary depicts the RoG plots of the rest of the proteins. We have to follow the minimum protein RoG value to pick a winner. We hypothesize that a protein with higher RoG is less stable, which is not the case for ligands given a small range of RoG. For example, the protein RoG of the highest RMSD complexes subset is 50.25 Å, 48.75 Å, and 49.25 Å (Figure 5d,e,f). The protein RoG values (Figure 5a,b,c) for the lowest RMSD complexes subset are 48.80 Å, 49.35 Å, 49.25 Å. Here the sequence is the same as observed in the lowest RMSD subset order for ligand. And we can perceive that the two of the complexes have the same value from high and low RMSD, but the 3$^{rd}$ complex has high RoG protein in case of the high RMSD sample set and 3$^{rd}$ complex have low RoG protein in case of the low RMSD sample set. This helps us pick the winner because the set of complexes with the average high RMSD also have high average protein RoG, and the group of low average RMSD also has low average protein RoG.

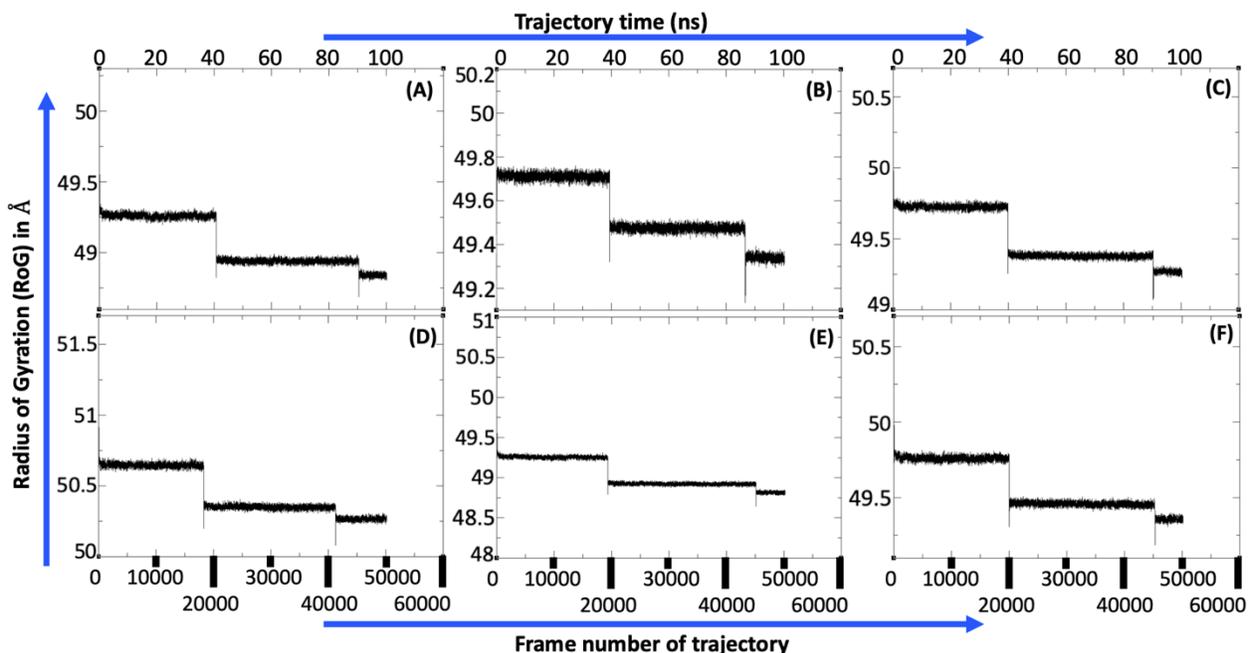

**Figure 5:** Radius of Gyration (RoG) of proteins' backbone having extreme RMSD. (A, B, C) Lowest RMSD sub-group: 12th, 13th, 18th ligands. (D, E, F) Highest RMSD sub-group: 5th, 11th, 14th ligands. The X-axis shows the frame number(bottom) & trajectory time(top) of the production run, and the y-axis is protein's RoG (Å).

*3.2.4 Solvent accessible surface area (SASA) analyses*

Solvent accessible surface area (SASA) measures how much fraction of the total surface area of the protein is accessible to the solvent. Since external molecules are integrated with solvent, the surface area exposure to solvent means exposure to external molecules. All SASA values are average SASA values observed over the entire MD production run of 100 ns. Usually, proteins have just high enough SASA to allow easy binding of a ligand/inhibitor but not higher, resulting in unstable protein. It means a fully unfolded ligand/protein along its backbone has the highest SASA among its conformations, albeit highest instability at the same time. The ligands with the lowest RMSD, i.e., 12$^{th}$, 13$^{th}$, 18$^{th}$, do not have the highest SASA among all the complexes (Figure 6). But they have more than average SASA values, i.e., 12$^{th}$, 13$^{th}$, & 18$^{th}$ ligands have 160 Å$^2$, 135 Å$^2$, 150 Å$^2$, respectively. Figure S6 in the supplementary depicts the SASA plots of the rest of the complexes. On the other hand, the ligands with the highest RMSD, i.e., 5$^{th}$, 11$^{th}$, 14$^{th}$, have the SASA as 70 Å$^2$, 130 Å$^2$, 160 Å$^2$, respectively. The second and third ligands for low- and high RMSD cases have almost the same SASA. The first ligand in the case of low RMSD ligands (12$^{th}$) has virtually double the SASA value than the first ligand of the high RMSD cases (5$^{th}$). Other complexes have even higher SASA values, i.e., the 5$^{th}$ complex (6m71_nsp12_7_8/ZINC000410177506/) has a SASA value of around 200 Å$^2$. This is a positive characteristic from the point of view of SASA. Still, we had to filter the ligands based upon a holistic approach with RMSD characteristics as a primary consideration. On the other hand, the 15$^{th}$ complex (7BV2_remdesivir/ZINC 000596164676) has the lowest average SASA of 10 Å$^2$. The native inhibitor Remdesivir was removed and replaced by the ligand under consideration in this complex.

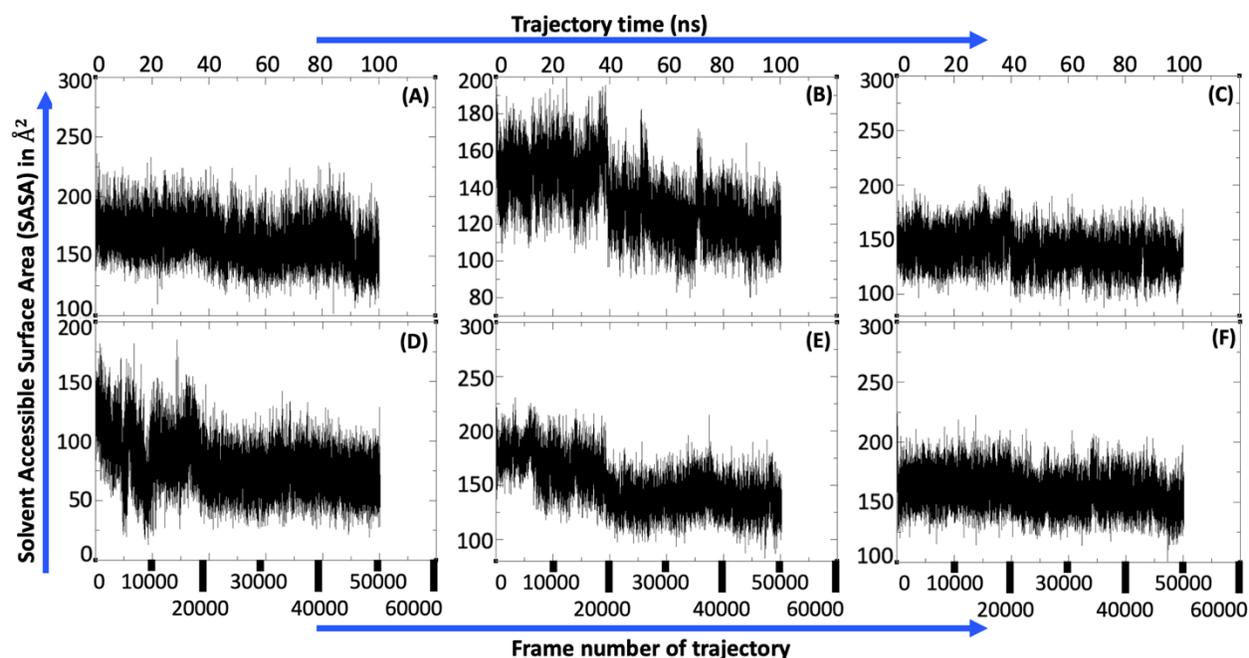

**Figure 6:** Solvent Accessible Surface Area (SASA) of ligands having extreme RMSD. (A, B, C) Lowest RMSD sub-group: 12th, 13th, 18th ligands. (D, E, F) Highest RMSD sub-group: 5th, 11th, 14th ligands The X-axis shows the frame number(bottom) & trajectory time(top) of the production run, and the y-axis is the protein's SASA (Å$^2$).

Finally, the protein's structural qualities, i.e., minimized structure's RMSD & MolProbity scores, are provided in Table S4. MolProbity and clash score of the proteins ranges from 91$^{st}$ percentile to 99$^{th}$ percentile, which supports the pipeline's integrity.

*3.2.5 Molecular Mechanics Poisson-Boltzmann Surface Area (MMPBSA) Analyses*

We calculated the ligand and protein binding free energy using the Molecular Mechanics Poisson-Boltzmann Surface Area (MMPBSA) approach implemented in AMBER20's MMPBSA.py subroutine script. Due to high data fluctuation given its stochastic nature, we compared the binding energy values for the six highlighted cases (3 from low RMSD[Table 3a] + 3 from high RMSD subset[Table 3b] as defined in the RMSD section). The binding energies for the three high RMSD cases $5^{th}$, $11^{th}$, & $14^{th}$ are -9.4409 kcal/mol, -0.5721 kcal/mol, & -9.8738 kcal/mol, respectively. The binding energies for three low RMSD cases $12^{th}$, $13^{th}$, $18^{th}$ are -6.5609 kcal/mol, -20.2606 kcal/mol, -4.5078 kcal/mol, respectively. We observed an almost direct correlation between the MMPBSA free binding energy and RMSD in the case of low and high RMSD ligand subsets. The ligands from the low RMSD ligand subset, on average, tend to have lower free energy than the average free binding energy of ligands from the high RMSD ligand subset. However, the $2^{nd}$ and $3^{rd}$ cases seem to be outliers in the case of high and low RMSD subsets, respectively. Binding free energy for the controlled complex of 7BV2/Remdesivir is calculated to be -20 eV.

**Table 3a:** Binding free energy from MMPBSA analysis for low RMSD ligand subset

| Ligand | 12 | 13 | 18 |
|---|---|---|---|
| Binding energy(eV) | -6.5609 | -20.2606 | -4.5078 |

**Table 3b:** Binding free energy from MMPBSA analysis for high RMSD ligand subset

| Ligand | 5 | 11 | 14 |
|---|---|---|---|
| Binding energy(eV) | -9.4409 | -0.5721 | -9.8738 |

### 3.3 Density Functional Theory (DFT) Analyses

*3.3.1 Molecular Electrostatic Map (MEP)*

We emphasize that all the ab-initio analyses were performed on systems containing only the ligands instead of the MD approach, which studied systems composing both proteins and ligands. From here onwards, we will primarily discuss connecting the ligand-only ab-initio results with the ligand/protein-complex MD results through various assumptions and hypotheses. In the MEP analysis, all six complexes appear to be covered with blue-iso-surfaces (Figure 7), indicating electron deficiency, i.e., bonded/unbonded interactions occur while the ligand gains an electron from the docked protein. This means the LUMO is more significant than the HOMOs in all the complexes to form bonded interaction and ultimately form hydrogen bonds and salt bridges. Figure S7 in the supplementary depicts the MEP plots of the rest of the complexes. In addition, we have performed Bader charge transfer analysis to probe charge dynamics further, as shown in the next section.

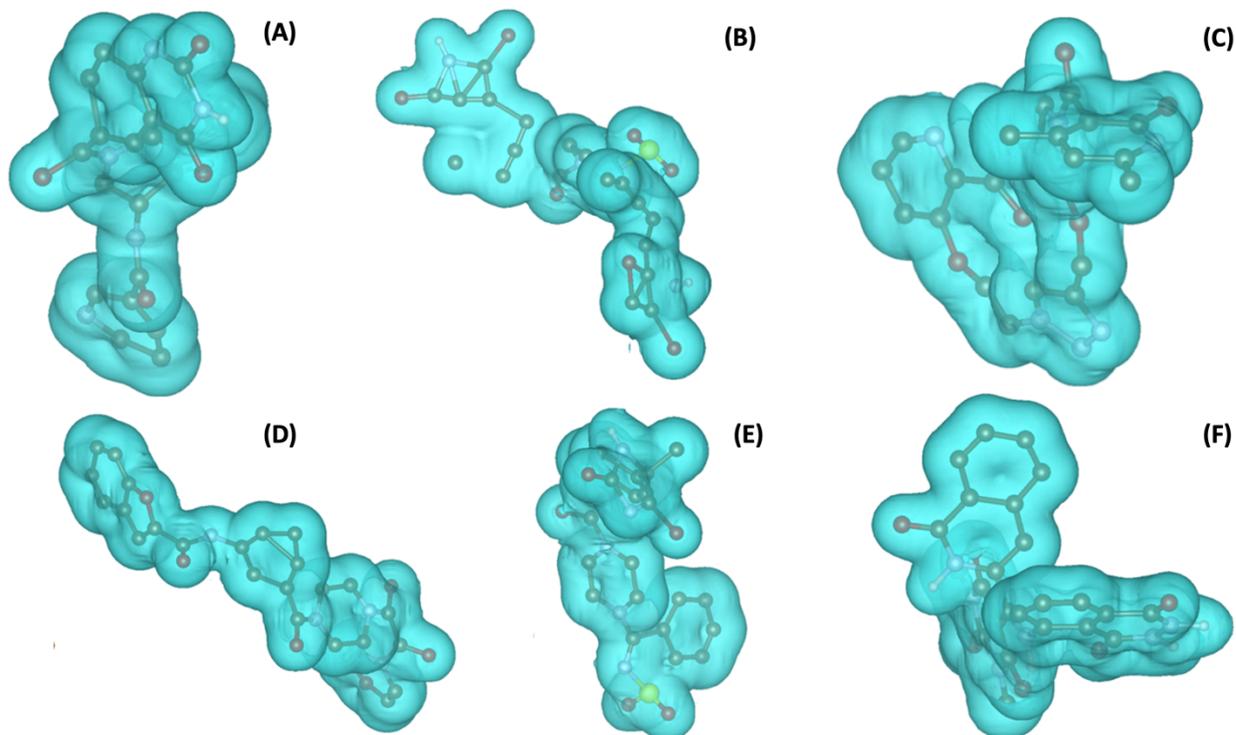

**Figure 7:** Molecular Electrostatic Potential (MEP) map of ligands having extreme RMSD. (A, B, C) Lowest RMSD sub-group: 12th, 13th, 18th ligands. (D, E, F) Highest RMSD sub-group: 5th, 11th, 14th ligands.

*3.3.2 Bader Charge Analysis*

We have further performed the Bader charge analysis to understand the charge potential of the individual atom/ion. We studied the same three complexes considered in the RMSD analysis having the least RMSD, i.e., 7bv2_nsp12-7/ZINC001176619532, 7bv2_nsp12-8/ZINC000517580540, and 7bv2_rna/ZINC000952855827. We notice the same observations in the Bader charge analysis (Figure 8) as in frontier molecular orbital analysis (Figure 10). The MEP plots denote that the ligand is more capable of accepting the electron than donating one. Consequently, one can see that the atoms residing in the volume overlapped by the LUMO (Figure 10) show negative Bader charge potentials as highlighted by black closed curves in Figure 8. This shows their receptivity of the electrons, resulting in bonded/unbonded interactions with the participating molecule/protein. Figure S8 in the supplementary depicts the Bader charge plots of the rest of the complexes' ligands. To further examine the extreme Bader charge transfer values, we have visualized the residues bonded with ligands through hydrogen bonds (Figure 9) and highlighted them with white closed curves. For the 12th ligand, the extreme Bader charge values (-1.13 eV to 1.64 eV) appear to arise from the interaction between the Oxygen & Nitrogen of the ALA467 & PHE 362 of protein and the Nitrogen & Oxygen of the top benzene ring of the ligand's naphthalene. A similar observation can be seen for HOMO & LUMO (Figure 10). We superimposed the closed curves of the Bader charge plots as that for HOMO/LUMO plots. We could do that because the Bader charge displays the extreme values precisely in those areas having the HOMO/LUMO iso-surfaces. That proves the accuracy of our Bader charge analysis approach since it almost matches with FMOs analysis. For this ligand, the LUMO orbital is concentrated around both benzene rings of the naphthalene because of the presence of the ALA467 & PHE 362 nearby.

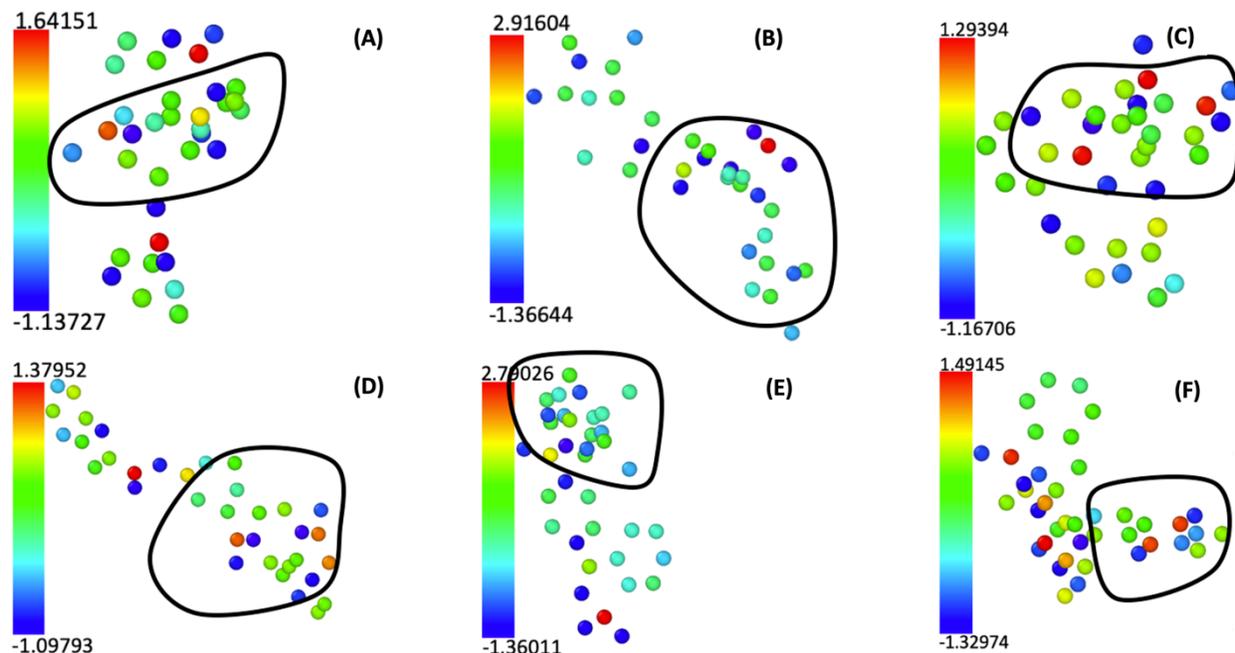

**Figure 8:** Bader charge transfer of ligands having extreme RMSD. (A, B, C) Lowest RMSD sub-group: 12th, 13th, 18th ligands. (D, E, F) Highest RMSD sub-group: 5th, 11th, 14th ligands.

For ligand 13th, the extreme values of the Bader charge analysis seem to be concentrated around volume containing S atom attached with two hydroxy functional groups and terminated with naphthalene on one end and Benzene heterocyclic ring on the other end. The benzene ring of naphthalene, further from the S atom, does have Oxygen and Nitrogen. The Nitrogen is balanced with an H atom and has a carbonyl group next to it. The other side of the S atom is terminated with a Benzene with 2 nitrogen atoms. These group S, hydroxy-functional group, carbonyl functionalized naphthalene interact with the nitrogen (ND2) and oxygen (OD1) atoms of residues ASN 436 and ASN 427. Out of which, oxygens are double bonded with the backbone of the residues. As shown in Figure 10b, most of the HOMO/LUMO volumes fall under the same closed curve highlighted on the Bader charge plot (Figure 8b). However, there appears to be LUMO covering the partial Benzene ring of hexahydroindan attached opposite to the Naphthalene terminal of the ligand. Finally, the 18th ligand is made primarily from 3 Benzene rings with a single heterocyclic Benzene ring attached as side chains through double bonds with terminal and middle Benzene rings of the backbone. Here most of the extreme Bader charge values (-1.16 to 1.29 eV) and HOMO/LUMO iso-surfaces appear to be concentrated over the region that contains the terminal Benzene that is not connected to the side chain benzene and the middle benzene ring. The terminal benzene ring has 1 nitrogen and is attached with 2 carbons through a single bond and 1 oxygen through a double bond. This ligand region appears to be the reason behind the formation of the most frequent hydrogen bond (27450 frames) with oxygen (OG1-HG1) of the residue THR 485 of the protein.

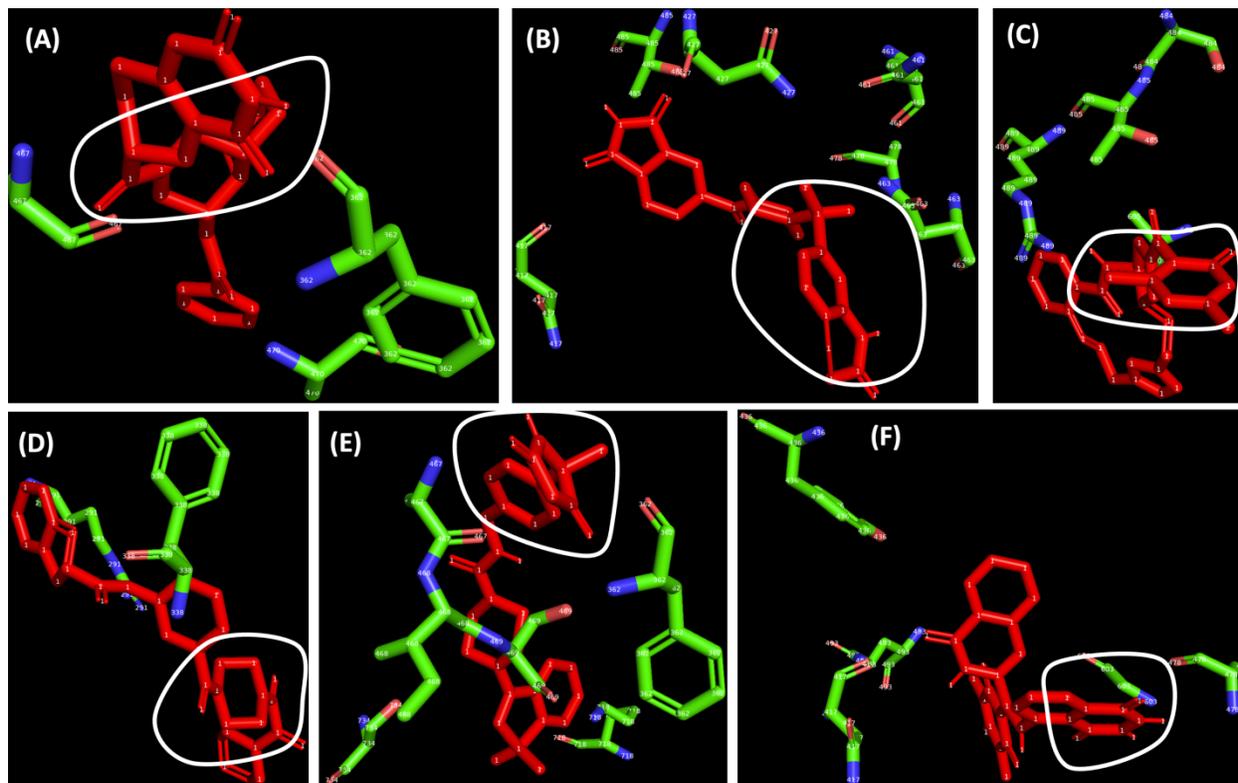

**Figure 9:** Stick representation of residues bonded with ligands through hydrogen bonds. (A, B, C) Lowest RMSD sub-group: 12th, 13th, 18th ligands. (D, E, F) Highest RMSD sub-group: 5th, 11th, 14th ligands. Ligand is represented in red.

On the other hand, the high RMSD ligand sub-group interacts with the protein binding site in such a way so that either the number of interactions are not enough or if they are large enough, then they are dispersed over the whole conformation of the ligand instead of being focused around a locale of the ligand, which displayed the iso-surfaces of HOMO/LUMO as shown in Figure 10. Ligand 5 constitutes one hexahydroindan heterocyclic ring with an oxygen atom in cyclopentane. And 1 benzene ring and 2 hetero-benzene rings, with one having 2 nitrogen atoms and the terminal benzene ring having a double-bonded oxygen atom, are attached through a nitrogen bond to the ligand's backbone. The conformation of the hetero-benzene rings makes them ideal candidates for the concentration of the HOMO/LUMO iso-surfaces. In this ligand, hydrogen bonds are formed only for two sets of residues and the ligand, i.e., ARG291@NH2–HH21⋯O3 and PHE338@N–H⋯O4. And these residues are not in close proximity to the hetero-benzene rings covered by HOMO/LUMO. This may be the prime reason this ligand falls under the category of high RMSD subset. Despite extreme values of the Bader charge transfer (-1.09 to 1.37 eV) in the same region covered by FMOs, there are some atoms of hexahydroindan, which exhibits high values for the Bader charge transfer because of the already cited hydrogen bonds. A similar phenomenon was observed for ligand 11, where a significant number of hydrogen bond interactions occur outside the region, overlapped by the HOMO/LUMO iso-surfaces. The compound had one hexahydroindan with a Sulphur atom in cyclopentane, 1 hetero-benzene, 1 benzene, and 1 hetero-Cyclopentane with 2 double-bonded attached oxygens. The majority of the high-frequency hydrogen bonds, i.e., 5634, 5268, 4671, 1361, 1235 frames, are formed between residue/ligand LYS718@NZ–HZ3⋯O1, –HZ2⋯O1, –HZ1⋯O1, –HZ2⋯O2, –HZ3⋯O2 respectively. They are further away from the LUMO/HOMO iso-surfaces among the hydrogen-bonded residues.

Furthermore, LIG835@N–H⋯ILE468@O is also a high-frequency bond and appears to be away from HOMO/LUMO overlapping volume, and so does the bond SER734@OG–HG ⋯O2. Conversely, the following bonds appear to be covered by the overlapping region of HOMO/LUMO iso-surfaces, i.e., LIG835@N4–HG ⋯ALA467@O, PHE362@N–H⋯O3, LIG835@N–H⋯SER469@OG. The Bader charge analysis agrees with the spatial distribution of the hydrogen bond formation. The highest positive value, i.e., 2.79 eV appears to be in the Sulphur atom interacting with residue LYS 718, and the highest negative value of -1.36 eV seems to be present in the atoms interacting with the residues LYS718, SER734, ILE 468, ALA 467. In ligand 14$^{th}$, the two sets of hetero-hexahydroindan are linked through the hetero-benzene side chain.  Out of which, one hexahydroindan contains benzene with 1 nitrogen & one double oxygen, and another hexahydroindan has benzene with 2 nitrogen & 2 double bonded oxygens. The side chain benzene ring contains 2 nitrogen connected with carbons and 3 double bonded oxygens.

We discuss the hydrogen bonds in the decreasing order of their frequency. At the top, we observed LIG835@N5–H3⋯GLY603@O, with 32130 frames, hydrogen bond formed between double-bonded hydrogen of GLY603 and one of the NH group of the terminal benzene ring. We believe this bond results from the highest Bader charge values (-1.32 to 1.49eV) obtained within the closed curve in Figure 8F. The second highest frequent bond is LIG835@N4–H2⋯ALA478@O with 23435 frames. This bond appears to be originated from one of the HOMO lobes that cover the hydrogen atom attached to benzene nitrogen, shown in the highlighted closed curve area in Figure 10F. The next hydrogen bond with 7510 frames is TYR436@OH–HH ⋯LIG@O3, which forms between benzene linked OH group of residue TYR 436 and double-bonded oxygen of non-terminal benzene of one of the hetero-hexahydroindan. And we observe the corresponding low Bader charge values(-1.32eV) on that double-bonded oxygen in Figure 8F. Followed by 3 hydrogen bonds GLN493@NE2–HE21⋯LIG@O2, LIG835@N2–H⋯ASN417@O, & ASN417@ND2–HD22⋯LIG@O1 with the frequency of 6593, 6132, & 2590 frames. The first of these three bonds are made between Nitrogen of residue GLN493 and oxygen of the side chain benzene ring. The 2$^{nd}$ and 3$^{rd}$ of these bonds are made between the same residue and benzene ring of the same hetero-hexahydroindan. Furthermore, the types of atoms involved are the same. The only difference is, in the 2$^{nd}$ bond, the donor is the ligand's nitrogen atom & the acceptor is the residue's oxygen atom, and this is reversed in the 3$^{rd}$ bond, i.e., the donor is residue's nitrogen atom, and acceptor is ligand's oxygen atom.

One can clearly observe a trend. For example, for a given ligand, if most high-frequency hydrogen bonds are observed, and corresponding extreme Bader charge transfer values occur in the region covered by the HOMO/LUMO iso-surfaces, it tends to belong to the low RMSD ligand subgroup.

And if the same majority do not fall under the region covered by the HOMO/LUMO iso-surfaces, it tends to belong to the high RMSD ligand subgroup.

### 3.3.3 *Frontier Molecular Orbitals (FMO)*

One of the primary causes of interaction in a biological system is hydrogen bond formation. A hydrogen bond is formed when the probability of an electron transfer is highest between a residue and ligand. We assume the probability of an electron transfer depends upon the following factors: the shape of the ligand relative to the binding cavity in the protein, the shape/overlapping of the frontier molecular orbitals (FMO), overlapping of the FMOs over the interlocking part and the volume of the FMOs. One can counter-argue here about our consideration of the ligand's shape and its implicit incorporation in FMOs. To refute it, we can consider that for a single docked pose of a ligand, there could be multiple shapes/volumes of the FMOs, and for a single volume/shape of the FMO, there could be various ligand poses. Therefore, the separate considerations of these two variables are better to characterize the ligand's dynamic in the binding site.

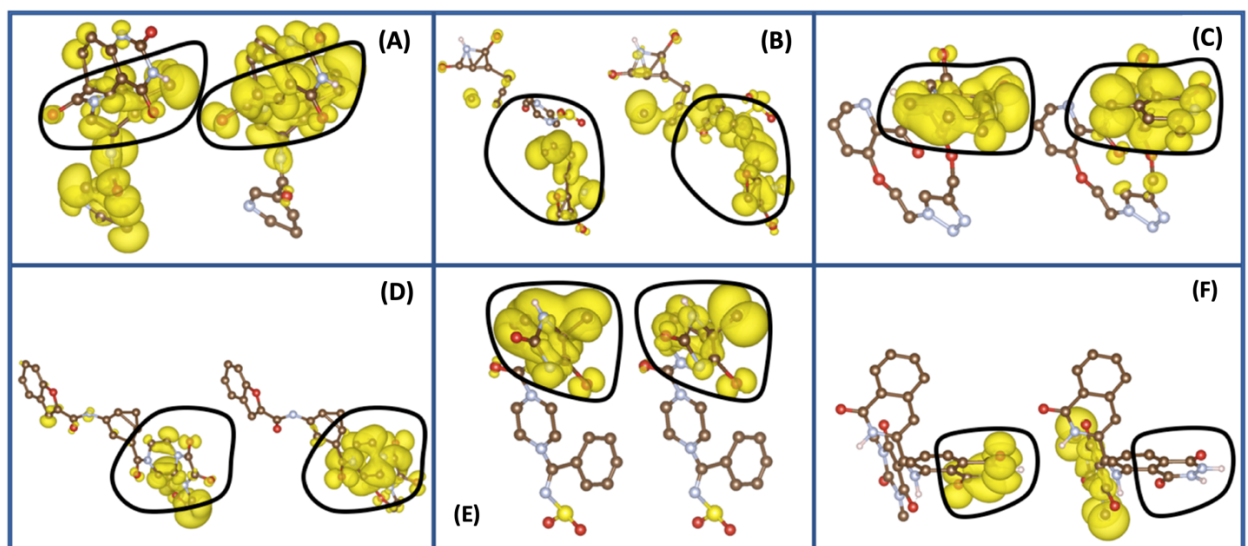

**Figure 10:** Frontier Molecular Orbitals (FMO) of ligands having extreme RMSD. (A, B, C) Lowest RMSD sub-group: 12th, 13th, 18th ligands. (D, E, F) Highest RMSD sub-group: 5th, 11th, 14th ligands. The left figure of each part shows the HOMO iso-surface, and the right figure shows the LUMO iso-surface.

We will be analyzing three complexes from high RMSD subgroup 6M71_nsp12_7_8/ZINC000410177506, 7BV2_nsp12-7/ZINC000616537204, 7BV2_nsp128/ZINC 001180048431, and three complexes from least RMSD ligand subgroup: 7bv2_nsp12-7/ZINC001176619532, 7bv2_nsp12-8/ZINC000517580540, 7bv2_rna/ZINC000952855827. Figure 10 shows the FMOs with 3D complex models for the previously mentioned six cases. Figure S9 in the supplementary depicts the FMO of the rest of the complexes. For the high RMSD ligand subset, the first complex has high RMSD because of the linear/non-interlocking geometry of the ligand in comparison to the binding pocket, despite having a significant FMO volume and their overlapping, as shown in Figure 10. The same holds for the second complex, despite having somewhat better inter-locking geometry of the ligand compared to the binding pocket, which is why this case has relatively higher RMSD than the first complex. And finally, the third case, even though it has an interlocking ligand profile, is not covered by either of the FMO orbitals. Moreover, the FMO orbitals do not overlap despite having decent volume.

On the other hand, in the case of the least RMSD ligands subset, the second complex has the least RMSD (0.45 Å) because of the better interlocking shape of the ligand with FMOs, covering that interlocking region in addition to overlapping orbitals despite having relatively lower orbitals volumes. Followed by second best, 3$^{rd}$ complex with RMSD (0.5 Å), which has the perfect combination of the FMOs overlapping each other and FMOs covering the whole ligand ring, which acts as an ideal interlocking geometry for a given binding site in a protein. Finally, the 1$^{st}$ case has the highest RMSD of 0.6 Å because of poor distribution/overlapping of the FMOs over the significant available interlocking area of the ligand & poor overlap of the FMOs despite having decent volume.

### 3.3.4 *Quantum Characteristics*

**Table 4:** Quantum chemical characteristics from ab-initio analysis for low & high RMSD ligand subset

| Ligand No. | E_gap(eV) | Ita(eV) | Omega(eV) | Sigma(eV$^{-1}$) | Zita(eV) |
|---|---|---|---|---|---|
| 5 | 1.3002 | .65010 | 29.7471 | 1.5382 | 6.2191 |
| 11 | .8070 | .40350 | 50.9861 | 2.4783 | 6.4145 |
| 14 | 1.1803 | .59015 | 25.0457 | 1.6944 | 5.4370 |
| Remdesivir (Benchmark case) | .9493 | .57465 | 36.8032 | 1.6691 | 5.2513 |
| 12 | .8585 | .42925 | 50.5960 | 2.3296 | 6.5906 |
| 13 | .8337 | .41685 | 38.4998 | 2.3989 | 5.6654 |
| 18 | 1.0814 | .54070 | 34.9425 | 1.8494 | 6.1471 |

We had measured the quantum chemical characteristics for all the 18 complexes as shown in Table S2 in the supplementary document. However, we will discuss only the subset of the three highest and three lowest RMSD ligands, as shown in Table 4. The characteristics measures are as follows: the gap between HOMO and LUMO orbital's energy (a measure of the ligand's reactivity and thus stability), Ita(hardness), omega(electrophilicity), Sigma (local softness), Zita(electronegativity). Given the variation in data, we discussed the characteristics for the whole subset as one entity and then compared them among high and low RMSD subcases. The gap between the HOMO and LUMO orbital's energy for the high RMSD sub-group (5$^{th}$, 11$^{th}$, 14$^{th}$ ligands) is as follows: 1.3002 eV, 0.8070 eV, 1.1803 eV and same characteristics for low RMSD sub-group (12$^{th}$ 13$^{th}$, 18$^{th}$) is 0.8585 eV, 0.8337 eV, 1.0814 eV respectively. We have observed a similar trend for hardness, i.e., the hardness tends to increase with the increase in RMSD of a molecule. For high RMSD subset the hardness measured is: 0.65010 eV, 0.40350 eV, 0.59015 eV and for low RMSD subset the hardness values are 0.42925 eV, 0.41685 eV, 0.54070 eV, respectively. We have observed the opposite trend in the case of electrophilicity. The electrophilicity decreases with the RMSD value of the ligand. The electrophilicity values of high RMSD group are 29.74 eV, 50.98 eV, 25.04 eV and as that for the low RMSD subgroup are 50.59 eV, 38.49 eV, 34.94 eV respectively. A similar trend was obtained in the case of local softness, i.e., high RMSD tends to lower the local softness. The local softness for high RMSD subgroup are 1.53 eV$^{-1}$, 2.47 eV$^{-1}$, 1.69 eV$^{-1}$ and for low RMSD group are 2.32 eV$^{-1}$, 2.39 eV$^{-1}$ and 1.84 eV$^{-1}$ respectively. We observed an inverse relationship between electronegativity and RMSD. The electronegativities of the high RMSD group are: 6.21 eV, 6.41 eV, 5.43 eV and for low RMSD subgroup are: 6.59 eV, 5.66 eV and 6.14 eV, respectively. We have performed the same set of calculations for the controlled study case of Remdesivir against 7BV2, and corresponding values are shown in Table 4. The three lowest RMSD ligands exhibit better quantum chemical characteristics than the controlled case of Remdesivir.

We can say from the collected data that the ligands with higher RMSD also tend to have high reactivity and lower stability. The electrophilicity helps us understand the reactivity, structural, & selectivity patterns in excited and ground states of the compounds. It highly correlates with the stabilization energy of the compound when it is saturated with electrons from docked virus protein enzyme. A high value of μ (chemical potential) and a low value of η (chemical hardness) indicate a better electrophile compound. Along with the HOMO-LUMO energy gap, the chemical hardness and chemical potential give important information about the reactivity of a compound. This combination also helps understand the charge transfer during the interaction between a compound and a protein. If a compound has a lower value of electrophilicity index, that means it has more probability of accepting electrons from its docked protein structure until it fills its frontier molecular orbitals. On the other hand, if a compound has a high electrophilicity index value, they are less likely to absorb an electron from its docked protein, and hence they are less chemically reactive. In the sample set of 18 complexes, the subset of 3 ligands with the least RMSD has more electrophilicity index than the other subset of 3 high RMSD ligands with low electrophilicity index. This means ligands with high reactivity have higher RMSD and vice versa. As usual, this is expected as if a more reactive compound is more unstable, it shall have higher RMSD than another compound with lesser reactivity.

### 3.4 Machine learning-based HOMO-LUMO energy gap

As a representative case, we selected the HOMO LUMO energy gap (HLEG) to be predicted with a Machine Learning (ML) model. Unfortunately, due to the limitations of the training data and computational resources, we could not capture the absolute values of the HLEG. Still, we successfully captured the qualitative trend observed in the ab-initio calculations of HLEG from three extreme cases of the highest RMSD and lowest RMSD ligands subsets.

**Table 5:** Quantum chemical characteristics from Machine learning approach for low & high RMSD ligand subset

| Ligand | Machine Learning(eV) | DFT (eV) | % RMS Error |
|---|---|---|---|
| 12 | 0.8013 | 0.8585 | 6.66 |
| 13 | 0.9307 | 0.8337 | 11.63 |
| 18 | 0.9711 | 1.0814 | 10.19 |
|  |  |  |  |
| 5 | 1.4083 | 1.3002 | 8.31 |
| 11 | 0.8578 | 0.8070 | 6.29 |
| 14 | 1.2874 | 1.1803 | 9.07 |

The qualitative trend obtained from ab-initio calculations is that the high RMSD sub-group of ligands, on average have high HLEG values, and the low RMSD sub-group of ligands, on average, have low HLEG values. This is the exact correlation obtained for the HLEG values obtained from the machine learning approach, as shown in Table 5. We believe the mismatch between the absolute values of HLG between ab-initio and machine learning approach can be attributed to one or any combination of the following: the difference between the source of the training data and production data, size of the training data, and the equation on which the model is based upon, i.e., Battaglia equations. In production codes, the first factor shall not exist because having the production data from the same source as the training data shall tend to eliminate the need to use machine learning in the first place. But it becomes a significant aspect when the training data set is limited, which is the issue in our case. This leads us to the second variable that the dataset QM9 has only 134,000 molecules. That is appropriate only for a limited set of scenarios. We believe this is the most significant parameter among all three behind missing the absolute values of HLEG. And lastly, the Battaglia equations are known to provide a good result for a beginner-level code, but the modification of the parameters is necessary to reduce the error, which is beyond the scope of this study. And we leave it for future work in these three variables of the machine learning analysis section of the article.

### 4. CONCLUSIONS

We studied more than 2 million unique protein-binding-site/ligand combinations, which were docked using high throughput virtual screening. These were filtered down to 90 complexes by selecting the top 10 binding energy complexes provided by the docking step for each of the nine proteins systems (binding site). They were further screened down to two ligands per binding site, resulting in 18 unique ligand/binding site complexes post ADMET analysis. Consequently, Molecular dynamics and first principle analysis were performed on these 18 unique complexes. Three complexes each for the lowest and highest ligand RMSD values were selected for the post MD simulation analysis based upon qualitative reasoning rationalized with MD and DFT computations. Ligands from the least RMSD subgroup are docked to the same protein (nsp12-7-8 complex) through three different binding sites (interface of nsp12/7, interface of NSP 12/8, &

binding site of template-primer RNA). The least RMSD sub-group is ligand 12th, 13th, 18th with RMSD as follows: 0.75Å, 0.4Å, 0.5Å and highest RMSD sub-group is ligand 5th, 11th, 14th with RMSD as 4Å, 2Å, and 1.9 Å. The hydrogen bond analysis demonstrates the low RMSD ligand sub-group have either higher frequency hydrogen bonds or more stable hydrogen bonds than their counterparts in the high RMSD ligand sub-group. The hydrogen bond-forming residues almost overlap with the corresponding RMSF plots of low and high RMSD ligand sub-group. We hypothesize that it is attributed to the location/volume of the iso-surfaces of FMOs and the extreme values of the charge transfer. We have observed an almost direct correlation between the extreme charge transfer values and hydrogen bonds.

Additionally, if the majority of the extreme charge values and thus hydrogen bonds are encapsulated by FMO's iso-surfaces, then that ligand tends to belong to the category of least RMSD ligand sub-group. And if the majority of extreme charge values and thus hydrogen bonds are not encapsulated by FMO's iso-surfaces, then that ligand tends to belong to the high RMSD ligand sub-group. For example, from the low RMSD ligand sub-group, in the 18th ligand, most of the extreme Bader charge values (-1.16 to 1.29 eV) and HOMO/LUMO iso-surfaces appear to be concentrated over the region that contains the terminal benzene ring. This ligand region seems to be the reason behind the formation of the highest frequent hydrogen bond (27450 frames) with oxygen (OG1-HG1) of the residue THR 485 of the protein. On the other hand, similarly, ligand 14th from high RMSD ligand sub-case, hydrogen bond LIG835@N5–H3···GLY603@O is most frequent with 32130 frames. This bond is formed between double-bonded hydrogen of GLY603 and one of the NH groups of the terminal benzene ring. We have further performed MMPBSA based binding energy analysis, which agrees with our RMSD observation that the low RMSD ligand subgroup has lower binding energy on average than the group with high RMSD, i.e., -10.4431 VS. -6.6289 eV. HOMO-LUMO energy gap from the ab-initio method is modeled to be 0.85 eV to 1.08 eV for low RMSD ligand subgroup compared to high RMSD ligand subgroup's range of 0.80 eV to 1.3 eV. Our results support the hypothesis that ligand needs to be less reactive to have low RMSD. And it agrees positively with other quantum characteristics. A GNN based machine learning model qualitatively supports these results. Our computational results provide insight into the drug repurposing for inhibition of different proteins of SARS-COV-2 and warrant further experimental investigations.

**SUPPORTING INFORMATION**

Additional information for all 18 ligands under study.

Figures: Docked models of ligand and protein; Root mean square deviation (RMSD) and fluctuation (RMSF); Radius of gyration (RoG); Solvent accessible surface area (SASA); Molecular Electrostatic Map (MEM); Bader charge analyses; Frontier Molecular Orbital (FMO); Stick representation of residues; Hydrogen bond analyses

Tables: Hydrogen bond analyses; Quantum chemical characteristics; Relationship between complex/ligand/protein number and complex/ligand/protein's identity

**AUTHOR INFORMATION**

**Corresponding Authors**

*Email : dibakar.datta@njit.edu

*Email : jk435@ddlab.edu

**Author Contributions**

JK and DD conceived the project. JK performed all computations and wrote the manuscript.

**Notes**

The authors declare no competing financial interest.

**ACKNOWLEDGEMENT**

The authors acknowledge the NJIT faculty start-up grant and the Extreme Science and Engineering Discovery Environment (XSEDE) for the computational facilities (Award Number – DMR180013).

**DATA AVAILABILITY**

The data reported in this paper is available from the corresponding author upon reasonable request.

**CODE AVAILABILITY**

The pre-and post-processing codes used in this paper are available from the corresponding author upon reasonable request. However, restrictions apply to the availability of the simulation codes, which were used under license for this study.